\documentclass[aps,prl,floatfix,twocolumn,superscriptaddress]{revtex4-1}
\usepackage{latexsym}
\usepackage{graphicx}
\usepackage{rotating}
\usepackage{hyperref}
\usepackage{amsmath,amssymb,amsfonts}
\usepackage{bm}
\usepackage{color}
\usepackage{times}

\newcommand{\ket}[1]{\ensuremath{| #1 \rangle}}

\newcommand{\dx}{$d_{x^2-y^2}$ }

\newcommand{\JJC}{ \exp\left( \sum_{i,j=1,N}{v^c_{|i-j|} n_i n_j}\right)}
\newcommand{\JJS}{ \exp\left( \sum_{i,j=1,N}{v^S_{|i-j|} S^z_i S^z_j}\right)}

\newcommand{\be}{\begin{eqnarray}}
\newcommand{\ee}{\end{eqnarray}}

\usepackage{natbib}
\usepackage[compact]{titlesec}
\titlespacing{\section}{0pt}{*0}{*0}
\titlespacing{\subsection}{0pt}{*0}{*0}
\titlespacing{\subsubsection}{0pt}{*0}{*0}
\begin{document}
\title{Phase Diagram of a Three Orbital Model for the high-T$_c$ cuprates}
\author{C\'edric Weber}
\affiliation{King's College London, Theory and Simulation of Condensed Matter (TSCM), The Strand, London WC2R 2LS}
\author{T. Giamarchi}
\affiliation{DPMC-MaNEP, University of Geneva, 24 Quai Ernest-Ansermet CH-1211 Geneva, Switzerland }
\author{C.M. Varma}
\affiliation{Department of Physics and Astronomy, University of California, Riverside, CA 92521}

\begin{abstract}
We study the phase diagram of an effective three orbital model of the cuprates using
Variational Monte-Carlo calculations (VMC) on asymptotically large lattices and exact diagonalization on a 24-site cluster. 
States with ordered orbital current loops (LC), itinerant Anti-ferromagnetism (AFM), d-wave superconductivity (SC), and the Fermi-liquid (FL) are investigated using appropriate Slater determinants refined by Jastrow functions for on-site and inter-site correlations. We find an LC state stable in
the thermodynamic limit for a range of parameters compatible with the Fermi surface of a typical hole doped superconductors provided the transfer integrals between the oxygen atoms have signs determined by the effects of indirect transfer through the Cu-4s orbitals as suggested by O.K. Andersen. The results of the calculations are that this phase gives way at lower dopings to an AFM phase and at larger copings to a SC phase followed by a FL phase.
\end{abstract}
\maketitle

Intense effort has been devoted to the phase diagram of the high Tc superconductors \cite{bonn_hightc_review}, especially the pseudo-gap phase. Possibilities suggested for the latter include RVB and/or preformed superconducting pairs \cite{kotliar_liu_dwave_slavebosons, Lee_rmp}, Loops of orbital current without broken translational symmetry (LC-phases \cite{varma_first_time_mf_theta2,varma_pseudogap_theory} and with broken translational symmetry \cite{chakravarty_ddw_pseudogap} and various other forms of lattice and magnetic order. The LC-phases are worth investigating in detail because of neutron observations 
\cite{fauque_neutrons_currents,martin_greven_currents_mercury} in four different families of cuprates of moments well compatible with the existence of such phases. 
Their on-set temperature is consistent with the the pseudogap temperature $T^*$ estimated from thermodynamic and transport measurements.
The fluctuations of such phases \cite{aji-cmv, aji-shekhter-cmv} could provide a path to also explain the properties of the strange metal phase \cite{mfl} as well as
the d-wave superconductivity. However several theoretical as well as experimental questions remain to be understood in relation to them.

In quasi-1D system, it was found in weak-coupling renormalization group calculations and numerical approaches that longer-range interactions or a multi-orbital nature of 
the unit cell are needed to stabilize the orbital current phases. \cite{orignac_2chain_long,schollwockbroken,chudzinski_ladder_long}. However in 1D these phases have a spatial 
modulation becoming incommensurate upon doping. 
Orbital current phases that do not break translational symmetry requires a multi-orbital model \cite{vsa,emery} that
includes the orbitals of the copper as well as the oxygens in the unit-cell. In two dimensions, a mean-field analysis \cite{varma_first_time_mf_theta2} of such a model showed the existence of the $\Theta$ phases when the Cu-O nearest-neighbor repulsion is strong enough. However, although the mean-field result is independently confirmed \cite{our_paper_orbital_currents}, going beyond mean-field, either with exact diagonalizations \cite{greiter_exactdiag_currents,greiter_exactdiag_currents2} or by 
variational Monte Carlo (VMC) calculations \cite{our_paper_orbital_currents} suggested an absence of currents for large lattice sizes for the canonical model of cuprates \cite{vsa, emery_3bands_model}.
VMC calculations suggested however that the main ingredient for the existence of such current was the frustration in kinetic energy. Such frustration was ensured in the mean-field approach by the conversion of the Cu-O interaction into an effective hopping. 

It is thus important to investigate more complete models in which alternative paths for the kinetic energy can provide such frustration. 
One such source of kinetic energy can be provided by apical oxygens \cite{our_paper_orbital_currents}. 
Another possible source is additional kinetic energy terms \cite{Andersen_electronic_structure_parameters_luca_nature}
that have been suggested in addition to the direct transfer between the oxygens. It was shown that the oxygen p$_{x,y}$ orbital has a much larger overlap
with the un-occupied copper 4s orbital than the direct O-O overlap, yielding on integration over the 4s orbital a different set of effective parameters, in which the effective O-O nearest neighbor transfer integral can have a sign opposite to that of the direct transfer
(for a derivation see the supplementary material \cite{supplementary_material}).
We find and will explain that this strongly affects the stability of the $\Theta_2$ phase.

In this paper we examine the role of such terms on the stability of orbital currents by performing a VMC investigation of the revisited three band Hubbard model of Ref.~\onlinecite{Andersen_electronic_structure_parameters_luca_nature}. We use
a Jastrow projected wave function that allows an unbiased investigation of the relative stability of a wide variety of phases including the $\Theta_2$ phase.
The key issues that we address in this work are:
(1) What is the range of model Hamiltonian parameters which supports the various phases; (2) Is this range of parameter consistent with the fermi-surface of the typical hole doped cuprate and with the properties of the Mott insulator-AFM half-filled phase.

Our work builds on Ref.~\onlinecite{our_paper_orbital_currents}, only a short summary of the method is therefore given.
The  three orbital model is defined by the Hamiltonian:
\begin{multline}
\label{eq:3band_hub}
  H =\sum_{(i,j)\sigma}{\left(t_{i,j} c^\dagger_{i\sigma}c_{j\sigma}+hc \right)}
  + \sum_{\alpha=p,d}{ U_{\alpha} \hat n_{\alpha \uparrow}\hat n_{\alpha \downarrow} }  \\ +
    \Delta \sum_{p,\sigma}{\hat n_{p\sigma}} + V_{dp} \sum_{d,p}{\hat n_d \hat n_p}
\end{multline}
where the sum over $(i,j)$ includes Cu-O and O-O neighbors,   $c$ stands for $p_{x,y}$ or \dx
orbitals depending on the site; $t_{i,j}$ are the transfer matrix elements
of magnitude $t_{dp}$ and respectively $t_{pp}$ for Cu-O and O-O matrix elements. Additional p$_x$-p$_x$ and p$_y$-p$_y$ transfer integrals,
which will be denoted by $t'_{pp}$ result from the large overlap between the p$_{x,y}$ and Cu-4s orbitals connect the oxygen neighbors of the same copper site.
$\Delta$, $U_d$, $U_p$ and $V_{dp}$ denote the charge transfer energy, the
on-site repulsions in the Cu-$d$ and O-$p$ orbitals, and the
nearest-neighbor repulsion between Cu-$d$ and O-$p$ orbitals.
A  set of parameters used typically for the cuprates \cite{emery_3bands_model,hybertsenrenormalization} is
$U_d=8$eV and $U_p=2$ eV, $|t_{dp}|=1.4$eV, $t_{pp}'=1$eV, and $V_{dp}=1$eV, although there is no direct determination of several of them.  We will investigate the dependence of the ground state properties with respect
to several of the parameters.

The variational wave function considered is built from the ground state $\Psi_0$ of the Hofstadter-like mean-field hamiltonian:
\begin{equation}
\label{eq:fullmf}
H^{MF} = \sum\limits_{(i,j)}{t_{ij}\ \chi_{ij}e^{i \theta_{ij}}\ c^\dagger_{i\sigma}c_{j\sigma}}
+ \Delta \sum_{p\sigma}{\hat n_{p\sigma}} + \sum_{i}{\bold{h_i}\bold{S_i}}
\end{equation}
where $\chi_{ij}$, $\theta_{ij}$, and $\bold{h_i}$ are variational parameters. The variables $\chi_{ij}$ and $\theta_{ij}$ are allowed to be different on each bond for a unit-cell.
$\theta_{ij} \ne 0$ is a requirement for time-reversal breaking through orbital currents; the geometry of flux within the unit-cell is given by closed loops of $\theta_{ij}$. The local magnetic field $\bold h_i$ allows antiferromagnetism.
Correlations and effects of quantum fluctuations of a variety of kinds  on the ground state wave-functions
are included by multiplying $\Psi_0$ by spin and charge Jastrow Factors,
\begin{equation*}
\mathcal{J}  =  \JJC \JJS
\end{equation*}
$v_{|i-j|}^c$ and $v_{|i-j|}^S$ are also variational parameters. $i=j$ in the charge Jastrow is equivalent to the local Gutzwiller projection.
We considered Jastrow factors with $|i-j|\leqslant 3$ Cu lattice spacings and checked that
they are negligible beyond this.
The minimization of the variational parameters is performed using a stochastic minimization procedure
\cite{umrigar_mcv_minimis,sorella_optimization_VMC} in which
the parameters of the uncorrelated part of the w.f. and
the Jastrow parameters are minimized at the same time.

\begin{figure}
  \begin{center}
    \includegraphics[width=1.0\columnwidth]{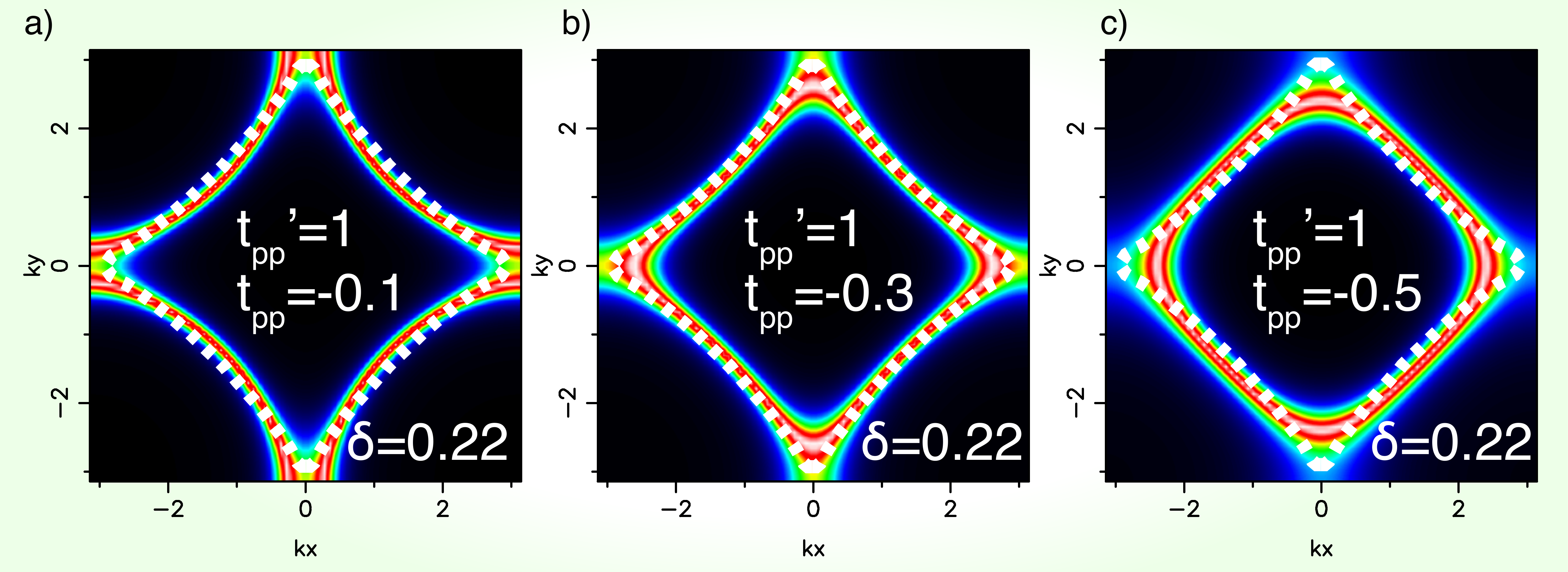}
    \caption{(Colors online) Fermi surface of the uncorrelated Hamiltonian (\ref{eq:3band_hub}) obtained for $\Delta=2$eV and a) $t_{pp}=-0.1$eV, b) $t_{pp}=-0.3$eV, c) $t_{pp}=-0.5$eV
    for $\delta=22\%$ hole doping. The Fermi surface curvature is weakly dependent on the charge transfer energy $\Delta$, but evolves significantly with the oxygen-oxygen
    transfer integral $t_{pp}$.
    The Fermi surface of LSCO extracted from ARPES data is shown for comparison \cite{spectral_function_LSCO_arpes_doping}.} \label{fig:arpes}
  \end{center}
\end{figure}
In the standard representation of the cuprates within a three band model,
the curvature of the Fermi surface (for a hole doped copper oxides) is given by the sign of $t_{pp}$. However, for the extended representation of
Ref. \onlinecite{Andersen_electronic_structure_parameters_luca_nature}, the two parameters $t_{pp}$ and $t_{pp}$' control the curvature
of the Fermi surface. The Fermi surface is in excellent agreement with the ARPES of LSCO for $t_{pp}'=1$eV and $t_{pp}=-0.3$eV (see (see Fig.~\ref{fig:arpes}.\textbf{b}).
These values are also consistent with their derivation \cite{supplementary_material} and the magnitude of the direct O-O transfer of about 0.7 eV typically used.
We also use $t_{pp}=-0.5$eV in the calculations in order to investigate the sensitivity of the results to this parameter.
Controlled calculations
\cite{millis_tpp_not_important} show that  the charge transfer gap does not depend much on $t_{pp}$ and $t_{pp}'$,
provided that the charge transfer energy is corrected $\Delta=\epsilon_p-\epsilon_d+2t_{pp}'$. Thus observable properties in the insulating state are unaffected by the new choice.

\begin{table}
\begin{center}
\begin{tabular}{|c|c|c|c|}
 \hline
Wave-function (w.f.)            &    E [eV]   & $\sigma$ [eV] & $J_{dp}$ [eV] \\
 \hline
FL               &   -1.700(1) &  0.064(1) &         0 \\
$\theta_2$    &   -1.996(1) & 0.0507(1) & 0.29(1)\\
$\theta_2$/J           &   -2.023(1) & 0.0409(2) & 0.22(2)\\
$\theta_2$/J/LS       &   -2.077(1) & 0.0498(1) & 0.22(1) \\
ED ($\bold{k}=(0,\pi),(\pi,0))$ & -2.1954(0) & 0 & 0  \\
ED ($\bold{k}=(0,0)$) & -2.1965(0) & 0 & 0 \\
   \hline
\end{tabular}
\caption{Variational energies $E$ and variance $\sigma$
of the different wavefunctions (w.f.) on an $8$-cell CuO$_2$ lattice with
$10$ holes and $S^z=0$ together with the exact ground state energies ED for total momentum $\bold{k}=(0,0)$ and $\bold{k}=(0,\pi),(\pi,0))$.
The variational Ansatz are: i) the Fermi sea projected with a local Gutzwiller projection or Fermi-liquid (FL), ii)  the mean-field orbital current w.f. in $\theta_2$ pattern projected with a local Gutzwiller projection, iii) $\theta_2$ optimized with
the Jastrow factors ($\theta_2$/J), iv) $\theta_2$/J improved by applying additionally one Lanczos step ($\theta_2$/J/LS).
All shown results are for $\Delta=0$ and $t_{pp}=-0.5$eV.}
\label{table1}
\end{center}
\end{table}

\begin{figure}
\begin{center}
    \includegraphics[width=0.5\columnwidth]{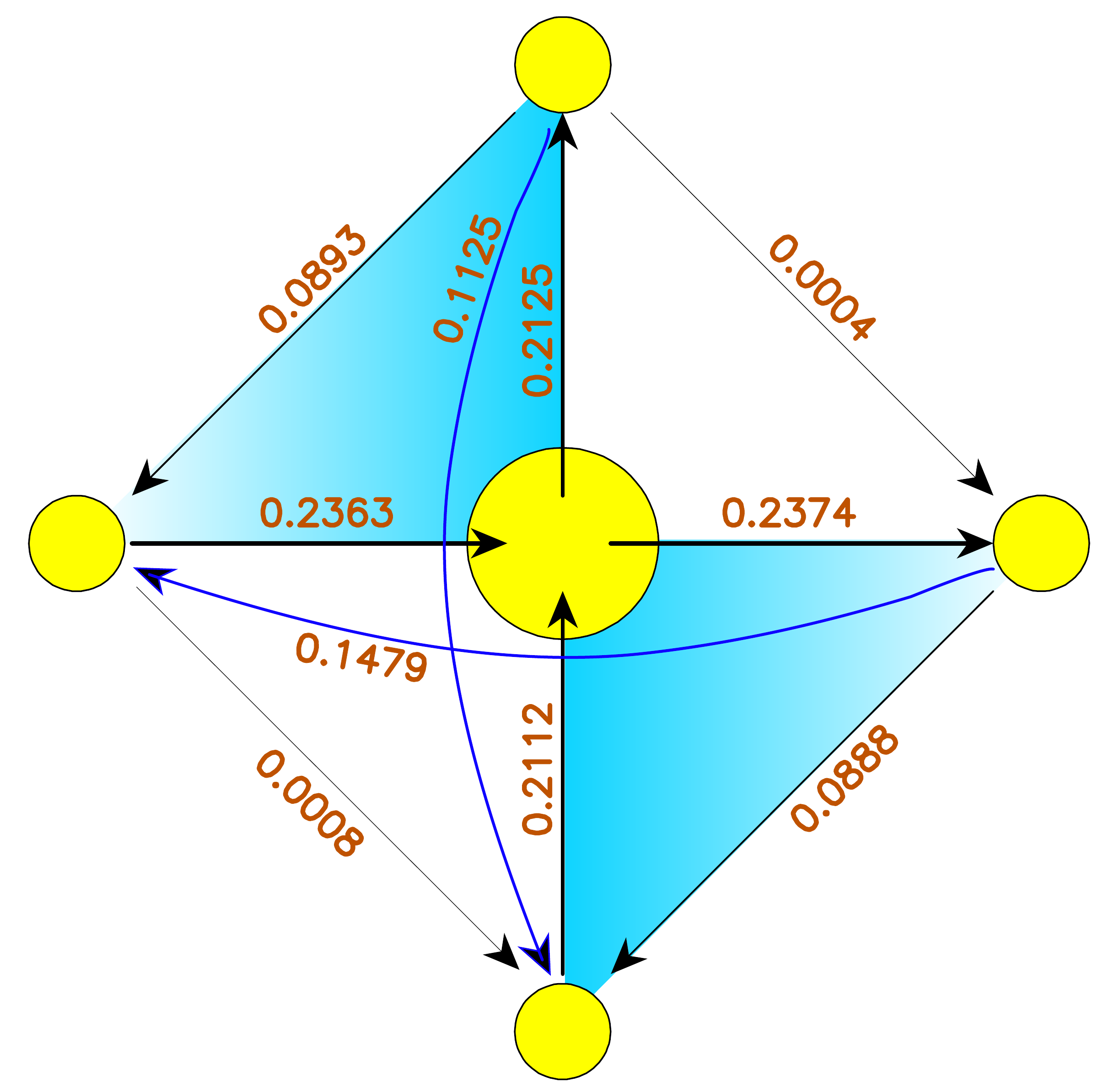}
    \caption{(Colors online) Pattern of current obtained from the variational w.f. $\theta_2$/J/LS on a $8$-copper lattice with
$10$ holes and $S^z=0$ (see Table~\ref{table1}). The current flows between orbitals lying on nearest neighbor sites (black arrow) and between
along the $p_x-p_x$ and $p_y-p_y$ oxygens orbitals (blue arrows). The current pattern has two circulating current loops (shaded areas) with
opposite chiralities.}
 \label{pattern}
  \end{center}
\end{figure}

We first discuss the low energy properties of a cluster of 8-CuO$_2$ cells and compare the variational results to the exact ground state energy in Table (\ref{table1}). As a reference, the energy of the Fermi-liquid (FL) state is shown.
We first allow the w.f. which allows any time reversal symmetry breaking pattern (the complex phases are optimized
on each of the link within a Cu-O$_2$ unit-cell). Remarkably, we find that the orbital currents are stabilized
and yield an effective energy optimization (see table~\ref{table1}, w.f. $\theta_2$). The
symmetry of the orbital current pattern (see Fig.~\ref{pattern}) consist of two orbital current loop, with
opposite chiralities, which is consistent with the theoretical proposal for the pseudo-gap phase of the cuprates
\cite{varma_first_time_mf_theta2}. 

The orbital current w.f. optimized with long-range Jastrow factors ($\theta_2$/J) and with a so-called \emph{Lanczos step}
\footnote{The Lanczos step is carried out by minimizing the w.f. in
an extended parameter space ($ \ket{ \psi } $ , $ \mathcal{H} \ket{ \psi }$).} 
($\theta_2$/J/LS) capture 95\% of the ground state energy (ED).
This suggests that the orbital current w.f. is a good candidate to describe the low energy physics of the three-band Hubbard model.
Remarkably, we also find that the degenerate exact eigenstates in the
$\bold{k}=(0,\pi)$ and $\bold{k}=(\pi,0))$ are only 0.0009eV ($\approx 10$ Kelvin) apart from the ground state. The presence of
very low energy states with finite momentum hint towards a possible orbital current instability
\footnote{Note that the orbital currents can not
be directly observed in the Lanczos calculations, since the ground state of a finite system does not break spontaneously any symmetry,
but remains in a superposition of quantum states.}.

Despite the good energy of our w.f., we notice that the obtained orbital current pattern for small lattices satisfies  conservation of the
current at each vertex to only within 10\% or less (but with overall current 0), as shown in Fig.~\ref{pattern}; this stems from the fact that the w.f. is not an eigenstate of the Hamiltonian (\ref{eq:3band_hub}), as discussed in Ref. \onlinecite{our_paper_orbital_currents}.

In order to understand the physics of the orbital currents, we compare the change in the different contributions to the variational energies. As expected
we find that the orbital current w.f. ($\theta_2$), without any further optimization such as the Jastrow or the Lanczos step, reduces the double
occupation, and reduces the local Coulomb energy from $E_U=0.42$eV (FL) down to $E_U=0.31$eV ($\theta_2$), and also
reduces the nearest-neighbour Coulomb repulsion from $E_V=0.54$eV (FL) down to $E_V=0.22$eV ($\theta_2$). This large potential energy
optimisation due to orbital currents is accompanied by increased kinetic energy; in particular the d-p kinetic energy is worsened from
$E_{d-p}=-3.63$eV down to $E_{d-p}=-0.76$eV which is in turn largely compensated by a reduction of
 O-O kinetic energy  from $E_{p-p}=0.97$eV down to $E_{p-p}=-1.77$eV.
Note that at low and moderate hole doping for $t_{pp}<0$, the signs of the oxygen-oxygen
overlaps in the Fermi liquid wave-function are such as to give a positive contribution to the kinetic energy due to the $t_{pp}$ term in the hamiltonian.
The orbital currents provide an efficient way to optimize both these and the interaction energy terms. This optimization is  ineffective both
at very low doping or for very large $\Delta$, where the $t_{pp}$ kinetic term is negligible due to the low oxygen hole densities.

The importance of the sign of $t_{pp}$ is that for chemical potential on the anti-bonding band of a three orbital model, as in the cuprates, the orbital current phase is favored if the product of the signs of the transfer integrals around the O-Cu-O triangles is positive. This is already suggested by the fact that in such a triangle in isolation, the two degenerate current carrying (complex) states lie in energy above the one real state even for the non-interacting model.

\begin{figure}
\begin{center}
    \includegraphics[width=1.0\columnwidth]{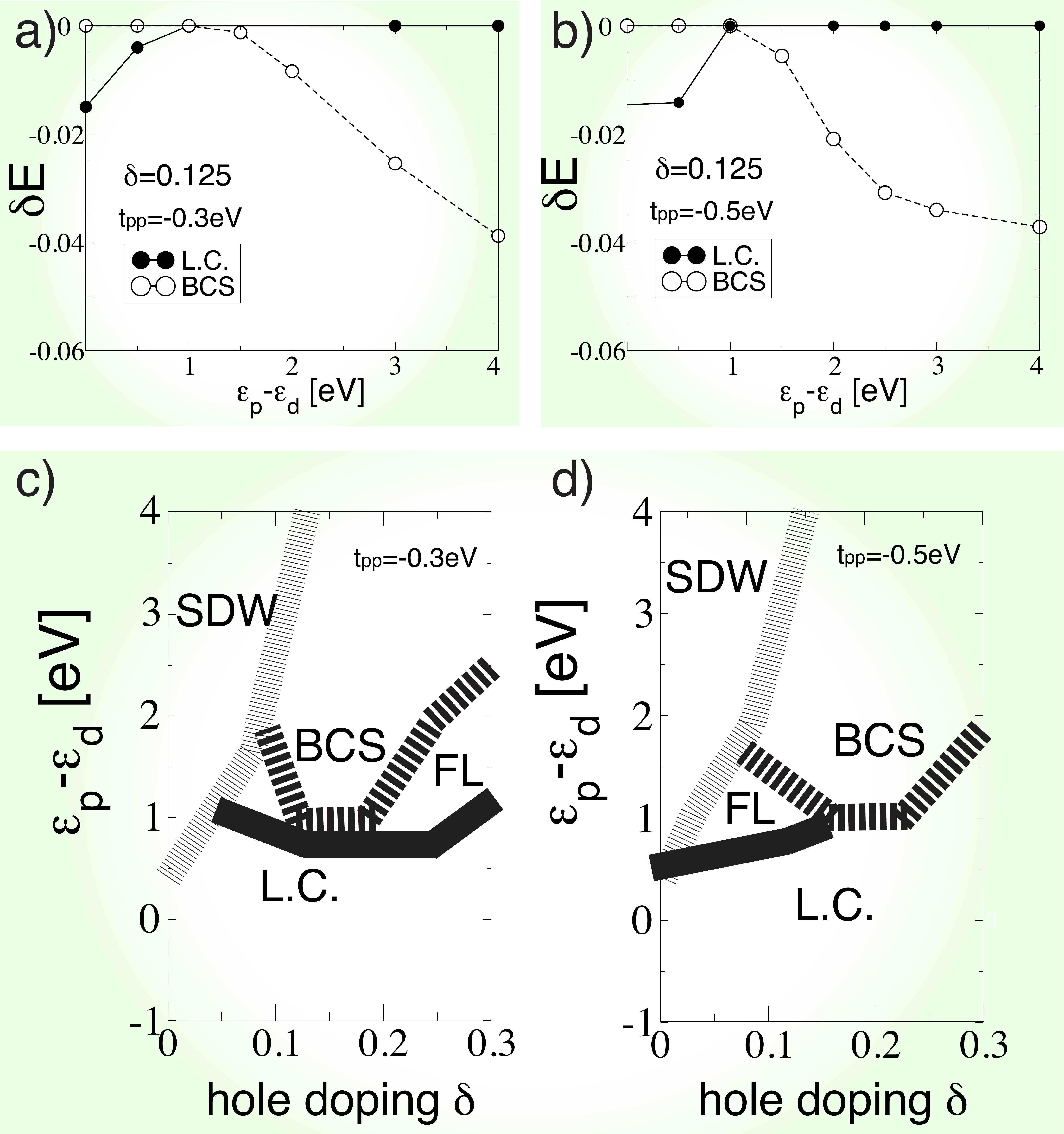}
    \caption{ (Colors online) Condensation Energy $\delta E$ of the loop current w.f. (filled circles) and of the d-wave superconducting wave-function (open circles) for 12.5\% hole doping
     for a) $t_{pp}=-0.3$ and b) $t_{pp}=-0.5$.  Region of parameters where the different instabilities have the lowest energies for
      c)  $t_{pp}=-0.3$ and d) $t_{pp}=-0.5$.
     Labels refer to the Loop-current phase (LC), to the antiferromagnetic phase (SDW),
     to the superconducting d-wave instability (BCS), and to the Fermi liquid (FL).
     The width of the phase boundaries comes from both the statistical error bar in the calculations 
     and from the discrete sampling of the phase space.
     Coexistence between the three different instabilities were not considered in this work.
     All calculations above are carried out for $N=192$.}
 \label{fig4}
  \end{center}
\end{figure}

Let us now consider the results for large clusters. We have performed calculations on lattices with $36$, $64$ and $100$ unit-cells sites, i.e.
$N=108$,$N=192$ and $N=300$ lattice sites. In order to avoid spurious finite size effect
induced by the artificial degeneracy of the variational wave-functions, we considered rotated geometries, with $T_1=(L,0)$ and $T_2=(1,L)$
lattice vectors, where $L=6,8,10$, and we used periodic boundary conditions in all cases. \footnote{For the specific case of the antiferromagnetic w.f. we have considered bipartite lattices with $T_1=(L,0)$ and $T_2=(0,L)$, see below in the text.}

In Fig.~\ref{fig4}.\textbf{a,b}, we show the condensation energy obtained for the orbital current $\theta_2$/J w.f., as a function of the charge transfer energy $\Delta$ at fixed doping $\delta \approx 12\%$ together with that for the d-wave superconducting phase.  The superconducting wave-function (see Fig.~\ref{fig4}.\textbf{a,b}) is obtained by replacing the Slater determinant with a
d-wave BCS wave-function and keeping the Jastrow factors.
We find that the loop current instability is present at small and moderate charge transfer energy $\Delta<1$eV. For large $\Delta$ the three band hamiltonian reduces to an effective one-orbital Hubbard or t-J model, and no orbital currents are found.
We find that at $\delta =12\%$ the d-wave BCS state is stabilized for $\Delta>1$eV (Fig.~\ref{fig4}.\textbf{a,b}).

At zero doping we find that the N\'eel magnetic long-range order is stabilized for $\Delta>1$eV, and
is stable to increasing doping when the charge transfer energy is increased, in agreement with early variational
Monte Carlo calculations done for the three band Hubbard model \cite{yanagisawa_vmc_3band}.

We summarize in Fig.~\ref{fig4}.\textbf{c,d} the phase diagram of the Fermi-liquid state and the states with the largest condensation energy for a given $\Delta$ and $\delta$. The antiferromagnetic phase is obtained by removing the degeneracy of up and down spins in the Slater determinant corresponding to a commensurate $(\pi,\pi)$ phase and with the Jastrow factors.  For $t_{pp}=-0.3$ and $\epsilon_p-\epsilon_d \approx 1$eV,  Fig.~\ref{fig4}.\textbf{c,d} shows that
the AFM state for $\delta < 5\% $, the LC state in a range starting at $\delta \approx 5\%$ to $\approx 12\%$, the d-wave state for $ 12\% <\delta < 20\%$, have the lowest Free-energy, followed by the one of the FL state. We have not considered coexistence of the latter instabilities (see e.g. Ref.~\onlinecite{giamarchi_tjhub}), but it is apparent from Fig.~\ref{fig4}.\textbf{c,d} that for reasonable parameters and given the general interaction free-energy $\gamma_{ij} |\Psi_i|^2 |\Psi_{j}|^2$ with $\Psi_{i}, \Psi_{j}$, the order parameters for AFM, LC, or SC states, and with $\gamma_{ij} > 0$ as expected, the Mott-AFM state at half filling gives way to an itinerant AFM state, followed by a co-existing LCO-SC state, then by a SC state alone and finally by a FL state, as $\delta$ is increased. This is fully compatible with the generic phase diagram of the cuprates. It should be clear that further tweaking of the parameters in a small range (for example reducing $t_{pp}$ by $< 10\%$) about the chosen parameters is likely to reproduce the small variations of the ground states of any of the cuprates with doping. 
\footnote{For example, for $t_{pp} = -0.35$ and other parameters remaining the same, the FL phase at low dopings in Fig.~\ref{fig4}.\textbf{c} disappears and the LC and BCS phase become degenerate within the error-bars of this calculation for $\Delta$ between 1 and 2 and $\delta$ between $5\%$ and $20\%$. We thank Lijun Zhu for these calculations.)}

\begin{figure}
\begin{center}
    \includegraphics[width=0.9\columnwidth]{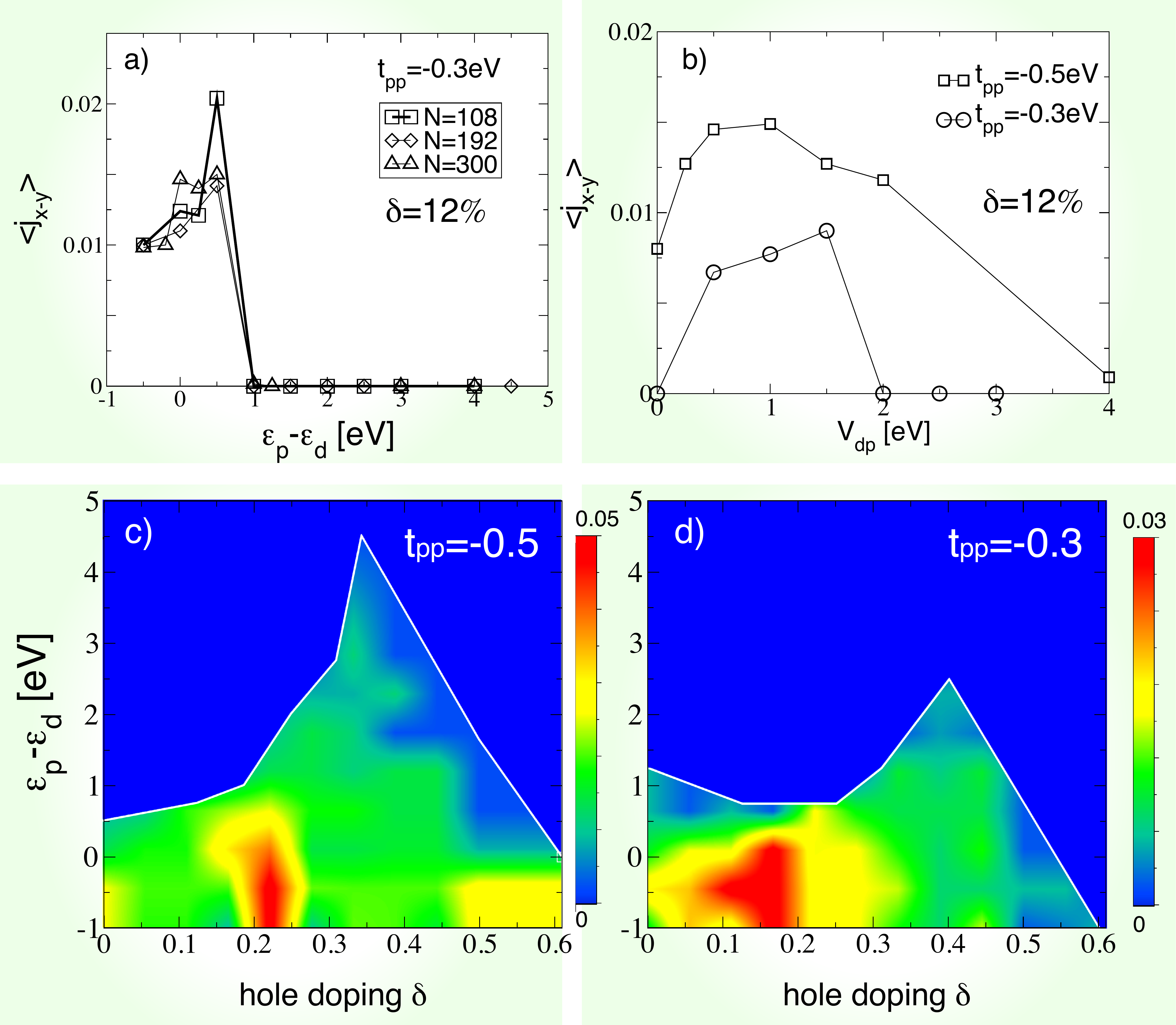}
    \caption{ (Colors online) Current amplitude on the $p_x$-$p_y$ link ($J_{x,y}$) obtained in the orbital current $\theta_2$/J w.f. as a function of a) the charge transfer energy $\epsilon_p-\epsilon_d$
     for different lattice sizes ($N$) at fixed hole doping $\delta \approx 12\%$, and as a function b) of the nearest neighbor Coulomb repulsion $V_{dp}$.
     Map of the current amplitude $J_{x,y}$ as a function of the hole doping $\delta$ and the charge transfer energy for c) $t_{pp}=-0.5$eV and d) $t_{pp}=-0.3$eV. All calculations above are carried out for $N=192$, and long-range Jastrow factors were considered for all w.f..}
 \label{thermo}
  \end{center}
\end{figure}

In Fig.~\ref{thermo}.\textbf{a}, we show the current amplitude, obtained on the $p_x$-$p_y$ link ($J_{x,y}$), and
measured in the orbital current $\theta_2$/J w.f., as a function of the charge transfer energy at fixed doping $\delta \approx 12\%$.
We do not observe significant variations with lattice size suggesting that the results
represent the thermodynamic limit. For large $\Delta$ the three band hamiltonian reduces to an effective one-orbital Hubbard or t-J model, as argued earlier \cite{ZhangRice,varma_first_time_mf_theta2} and no orbital currents are found. We note that a moderate Coulomb repulsion $V_{dp}=1$eV reinforces the current amplitudes when compared to $V_{dp}=0$eV (Fig.~\ref{thermo}.\textbf{b}), although unrealistic large Coulomb repulsions $V_{dp}>3$eV tend to suppress the currents.
The latter can be understood by the fact that increasing $V_{dp}$ leads to an effective increase of the charge transfer energy (Hartree shift).
Indeed, for large charge transfer energy $\Delta>4$eV we do not find any signature of the loop current phase.
In this range, our results are quite similar to earlier calculations \cite{tremblay_competition_sdw_bcs,tremblay_one_band_competition_antiferro,ivanovantiferromagnetism} of the phase diagram of the Hubbard model which show a large doping region of itinerant anti ferromagnetism, followed by co-existence with d-wave superconductivity and the superconductivity alone.

We extended the calculations to the large doping range $\delta<60\%$ (see Fig.~\ref{thermo}.\textbf{c,d}).
We find that the orbital currents are stable across most of the doping range when $\Delta \approx 0$.
For $\Delta <-1$ (not shown) we obtain that the current are suppressed. We note that in this limit
the holes fill the oxygens rather than the copper orbitals, which is an unphysical limit for copper oxides.
For larger charge transfer energies and $t_{pp}=-0.5$eV (Fig.~\ref{thermo}.\textbf{c}),
the amplitude of the loop current phase varies with doping being largest in the range $20-25\%$ and decreasing strongly for larger doping.
Nevertheless, at very large doping $ \delta > 30\% $, we find that the loop current phase is stable,
albeit with very weak current amplitudes. The latter region is in our view
probably not physical, since in the region of high doping other bands might be important \cite{luca_xas_edge_state,our_paper_apical}.

Very interestingly, the loop-currents are largest in small region of parameters $\Delta \approx 0$ and respectively $\delta \approx 0.2-0.25$
and $\delta \approx 0.1-0.15$  for $t_{pp}=-0.5$eV (Fig.~\ref{thermo}.\textbf{c}) and $t_{pp}=-0.3$eV (Fig.~\ref{thermo}.\textbf{d}).
which is close to the overdoped region of the superconducting phase of hole doped cuprates, which suggests that the
loop current phase is a candidate for the quantum critical point in this particular region of parameters.

To summarize, we have carried out a detailed study of several broken symmetry phases of an effective three orbital model
for the cuprates, with the special new feature that it includes Cu-4s mediated oxygen-oxygen transfer, as suggested by O.K. Andersen.
This indirect O-O hybridization of Andersen leads to an ambiguity regarding the sign of the $t_{pp}$ transfer integral. We have made a choice consistent with the direct $t_{pp}$ and in agreement with the Fermi surface of a typical hole doped superconductor. We show that this new effective model for cuprates yields a stable LC  phase at finite doping, consistent with the phase diagram suggested earlier and discovered by neutron scattering.
We extended the calculations to large clusters by VMC, and validate our theory by deducing a map of
the amplitude of the orbital currents as a function of the doping and the charge transfer energy, as well as the correct sequence of AFM, LC, SC and FL states as a function of doping observed generically in cuprates.

We thank Lijun Zhu, O.K Andersen, G. Kotliar, P. Bourges and M. Greven for many illuminating discussions.
This work was supported in part by the Swiss NSF under MaNEP and Division II. CMV's work was supported by NSF grant DMR-1206298.

\bibliographystyle{unsrtcedric}

\bibliography{/Volumes/BigLionImage/Users/cweber/Documents_Cedric/BIBLIO/biblio_cedric}

\end{document}


\title{Phase Diagram of a Three Orbital Model for the high-T$_c$ cuprates \\ Supplementary material}
\author{C\'edric Weber}
\affiliation{King's College London, Theory and Simulation of Condensed Matter (TSCM), The Strand, London WC2R 2LS}
\author{Thierry Giamarchi}
\affiliation{DPMC-MaNEP, University of Geneva, 24 Quai Ernest-Ansermet CH-1211 Geneva, Switzerland }
\author{Chandra M. Varma}
\affiliation{Department of Physics and Astronomy, University of California, Riverside, CA 92521}

\maketitle

\section{ Effective indirect transfer integrals through Cu-4s.}

\begin{figure}
 \includegraphics[width=1.0\columnwidth]{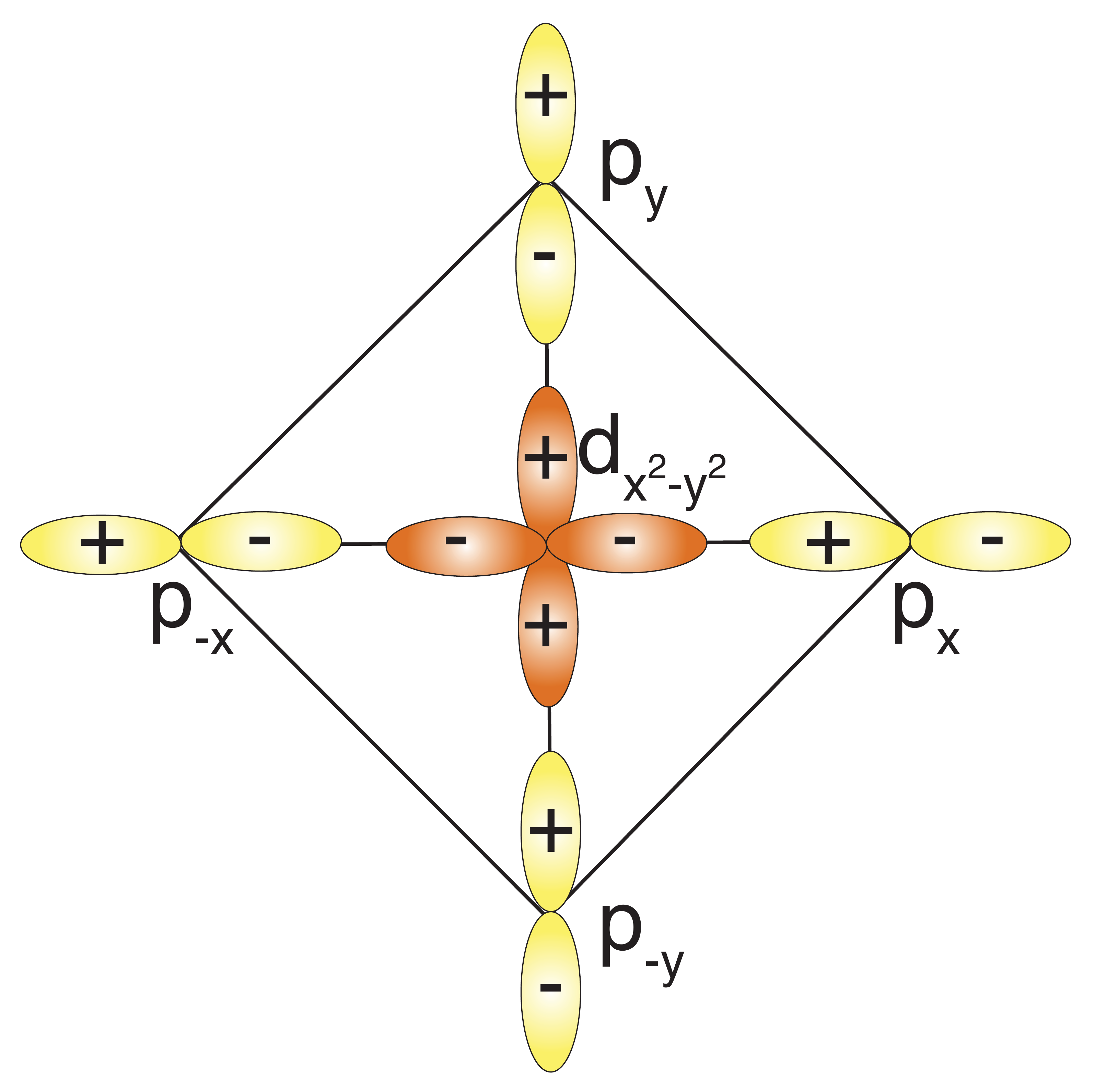}
 \caption{ The Cu and O -ions in a unit-cell with a particular choice of gauges on the p$_x$, p$_y$ and d$_{x^2-y^2}$ orbitals.}
 \label{fig5}
\end{figure}

Consider an O on the x-side of a Cu ion and one on the y-side of the same Cu ion. The direct hopping integral between the oxygens is $t^d_{pp}$. 
There is also an important indirect hopping \cite{Andersen_electronic_structure_parameters_luca_nature} between the same two O ions through the 4s state of the Cu. Let the hopping integrals between the Cu-4s and the O p-states be $t_{px,s}$ and $t_{py,s}$. It is easy to see (see Fig.~\ref{fig5}) that for any gauge choice:
\be
sgn(t_{px,s}t_{py,s}) = sgn(t^d_{pp}).
\ee

Let the energy of the oxygen ionic states be $\epsilon_p$ and the Cu- 4s state be $\epsilon_s$ with $\epsilon_s > \epsilon_p$. 
The indirect hopping between the two O orbitals through the high energy states is:
\be
t^i_{px,py} = 2\frac{t_{px,s}t_{py,s}}{\epsilon_p-\epsilon_s}
\ee
Therefore 
\be
sgn(t^i_{px,py}) = - sgn(t^d_{pp})
\ee
for any pair of nearest neighbor O ions.\\

Moreover, we obtain from the same procedure that the self-energy of a given O p-orbital is:
\be
\delta \epsilon_p = \frac{|t_{ps,x}|^2}{\epsilon_p-\epsilon_s}
\ee
and thus is negative, which confirms that the procedure is consistent.

Since $|t_{px,s}t_{py,x}|$ is expected \cite{Andersen_electronic_structure_parameters_luca_nature} to be about an order of magnitude
larger than $|t^d_{pp}|^2$ and $\epsilon_s - \epsilon_p$ is also an order of magnitude larger than $t^d_{pp}$, we should expect $|t^i_{px,py}| \gtrsim  |t^d_{pp}|$. 
The net effective $t_{pp}$ is the sum of the latter two. \\

We must also estimate the indirect O-O hopping integral across the Cu:
\be
t'_{pp} = 2\frac{t_{p+x,s}t_{p-x,s}}{\epsilon_p-\epsilon_s}.
\ee
 The $sgn(t_{p+x,s}t_{p-x,s})$ is negative. So, this gives $t^i_{p+x,p-x} >0$ of similar magnitude as $|t^i_{p,p}|$.
 
 Given the estimates above and the fact that $t^d_{pp}$ has been estimated to be about 0.7 eV, we have  used in our work
  $t'_{pp} \approx 1eV$ and $t_{pp} \approx -0.3 eV$ consistently. This choice is also motivated by the fact that these 
  values give a fermi-surface consistent with experiments done for a typical hole doped copper oxides (LSCO).

\bibliographystyle{unsrtcedric}

\bibliography{/Volumes/BigLionImage/Users/cweber/Documents_Cedric/BIBLIO/biblio_cedric}